\documentstyle[12pt,epsfig]{article}
\def\mathrm{\rm}

\advance\textheight by \topskip
\begin{document}
\tolerance=100000
\thispagestyle{empty}
\setcounter{page}{0}

\def\ie{\hbox{\it i.e.}{}}      \def\etc{\hbox{\it etc.}{}}
\def\eg{\hbox{\it e.g.}{}}      \def\cf{\hbox{\it cf.}{}}
\def\etal{\hbox{\it et al.}}    \def\vs{\hbox{\it vs.}{}}
\def\dash{\hbox{---}}
\newcommand\refq[1]{$\cite{#1}$}
\def\lq{{\rm LQ}}
\def\fb{~{\rm fb}}
\def\pb{~{\rm pb}}
\def\ev{\,{\rm eV}}
\def\mev{\,{\rm MeV}}
\def\gev{\,{\rm GeV}}
\def\tev{\,{\rm TeV}}
\def\wh{\widehat}
\def\wt{\widetilde}
\def\abs#1{\left| #1\right|}   
\newcommand{\slsh}{\rlap{$\;\!\!\not$}}     
\newcommand{\as}{{\ifmmode \alpha_S \else $\alpha_S$ \fi}}
\newcommand{\ep}{\epsilon}
\newcommand{\be}{\begin{equation}}
\newcommand{\ee}{\end{equation}}
\newcommand{\bea}{\begin{eqnarray}}
\newcommand{\eea}{\end{eqnarray}}
\newcommand{\ba}{\begin{array}}
\newcommand{\ea}{\end{array}}
\newcommand{\bi}{\begin{itemize}}
\newcommand{\ei}{\end{itemize}}
\newcommand{\bn}{\begin{enumerate}}
\newcommand{\en}{\end{enumerate}}
\newcommand{\bc}{\begin{center}}
\newcommand{\ec}{\end{center}}
\newcommand{\ul}{\underline}
\newcommand{\ol}{\overline}
\newcommand{\sm}{${\cal {SM}}$}
\newcommand{\mssm}{${\cal {MSSM}}$}
\newcommand{\Dir}{\kern -6.4pt\Big{/}}
\newcommand{\Dirin}{\kern -10.4pt\Big{/}\kern 4.4pt}
\newcommand{\DDir}{\kern -7.6pt\Big{/}}
\newcommand{\DGir}{\kern -6.0pt\Big{/}}

\def\Ord{\buildrel{\scriptscriptstyle <}\over{\scriptscriptstyle\sim}}
\def\OOrd{\buildrel{\scriptscriptstyle >}\over{\scriptscriptstyle\sim}}

\relax
\def\ap#1#2#3{
        {\it Ann. Phys. (NY) }{\bf #1} (19#3) #2}
\def\app#1#2#3{
        {\it Acta Phys. Pol. }{\bf #1} (19#3) #2}
\def\cmp#1#2#3{
        {\it Commun. Math. Phys. }{\bf #1} (19#3) #2}
\def\cpc#1#2#3{
        {\it Comput. Phys. Commun. }{\bf #1} (19#3) #2}
\def\ijmp#1#2#3{
        {\it Int .J. Mod. Phys. }{\bf #1} (19#3) #2}
\def\ibid#1#2#3{
        {\it ibid }{\bf #1} (19#3) #2}
\def\jmp#1#2#3{
        {\it J. Math. Phys. }{\bf #1} (19#3) #2}
\def\jetp#1#2#3{
        {\it JETP Sov. Phys. }{\bf #1} (19#3) #2}
\def\ib#1#2#3{
        {\it ibid. }{\bf #1} (19#3) #2}
\def\mpl#1#2#3{
        {\it Mod. Phys. Lett. }{\bf #1} (19#3) #2}
\def\nat#1#2#3{
        {\it Nature (London) }{\bf #1} (19#3) #2}
\def\np#1#2#3{
        {\it Nucl. Phys. }{\bf #1} (19#3) #2}
\def\npsup#1#2#3{
        {\it Nucl. Phys. Proc. Sup. }{\bf #1} (19#3) #2}
\def\pl#1#2#3{
        {\it Phys. Lett. }{\bf #1} (19#3) #2}
\def\pr#1#2#3{
        {\it Phys. Rev. }{\bf #1} (19#3) #2}
\def\prep#1#2#3{
        {\it Phys. Rep. }{\bf #1} (19#3) #2}
\def\prl#1#2#3{
        {\it Phys. Rev. Lett. }{\bf #1} (19#3) #2}
\def\physica#1#2#3{
        { Physica }{\bf #1} (19#3) #2}
\def\rmp#1#2#3{
        {\it Rev. Mod. Phys. }{\bf #1} (19#3) #2}
\def\sj#1#2#3{
        {\it Sov. J. Nucl. Phys. }{\bf #1} (19#3) #2}
\def\zp#1#2#3{
        {\it Z. Phys. }{\bf #1} (19#3) #2}
\def\tmf#1#2#3{
        {\it Theor. Math. Phys. }{\bf #1} (19#3) #2}

\def\preprint{{\it preprint}}

\begin{flushright}
{\large DTP/97/16}\\
{\large ETH--TH/97--10}\\
{\rm March 1997\hspace*{.5 truecm}}\\
\end{flushright}

\vspace*{\fill}

\begin{center}
{\Large \bf QCD Corrections and
 the Leptoquark Interpretation of the HERA High--$Q^2$ Events
\footnote{E-mails:
 kunszt@itp.phys.ethz.ch;
 W.J.Stirling@durham.ac.uk.} }\\[0.5cm]
{\large 
Z.~Kunszt$^{a}$ and W.~J.~Stirling$^{b,c}$}\\[0.15 cm]
{\it a) Institute of Theoretical Physics, ETH, Z\"urich, Switzerland.}\\[0.15cm]
{\it b) Department of Physics, University of Durham,}\\
{\it c) Department of Mathematical Sciences, University of Durham,}\\
{\it South Road, Durham DH1 3LE, United Kingdom.}\\[0.15cm]
\end{center}
\vspace*{\fill}
\begin{abstract}
{\noindent\small 
The excess of high--$Q^2$ events found by H1 and ZEUS
at HERA in $e^+p$ deep-inelastic scattering above the Standard Model
prediction 
motivates us to calculate
 the NLO QCD corrections to the HERA scalar leptoquark (or squark)  production
cross sections. We find that the corrections are significant, of order
50\% in the mass range of interest. We also calculate the leptoquark
average transverse momentum squared and find it to be rather small.
Various leptoquark production cross sections at
the Tevatron $p \bar p$ collider are also considered.
We investigate in detail  the leptoquark interpretation of the
HERA data. First we assume a  minimal leptoquark model with a single diagonal
Yukawa coupling to first family lepton and quark mass eigenstates only.
In this case constraints
from atomic parity violating experiments
allow only  isodoublet scalar leptoquark production at HERA.
This interpretation can  be confirmed
or ruled out in the near future by
high luminosity data at the  Tevatron.
The Tevatron data  already appear to rule out the
vector leptoquark interpretation of the HERA data.
We also consider  a more general  model
which allows for all possible left-handed,  right-handed,
 flavour and lepton number changing couplings.
The allowed values of the  Yukawa couplings of this general model
offer several different interpretations of the data which are
 radically different from the minimal model solutions.  
However these somewhat {\it ad hoc} tuned solutions  can  easily
be tested by future HERA experiments.}
\end{abstract}
\vspace*{\fill}

\newpage
\section{Introduction}
Recently the HERA collaborations H1  \cite{H1HQ2} and
ZEUS  \cite{ZeusHQ2} have presented new results on
high--$Q^2$ deep inelastic $e^+p$ scattering based on   $14\, \pb^{-1}$
and $20\,\pb^{-1}$ luminosity data, respectively. The data indicate an
excess of events with respect to the Standard Model expectation
predicted by the
conventional next-to-leading order QCD--improved parton model in a new
kinematical range unexplored by previous experiments.
At high $Q^2$ values $Q^2> 15000\gev^2$ H1 finds
  7 excess events  while ZEUS finds 5 events. An interesting
feature of the H1 events is that they are  clustered in a single bin of
the  invariant mass
of the jet--lepton system ($M_{\rm ej}$ = $\sqrt{xs}$) around $200 \gev$
with an estimated experimental resolution of $6\gev$ \cite{H1HQ2}.
The ZEUS events do not exhibit such a clustering,  but this
is not in conflict with the H1 data since ZEUS have a larger $M_{\rm ej}$
resolution.
H1  have also published results on charged current
events \cite{H1HQ2}. Although they find more events than expected, the
excess is not statistically  significant.
If the signal is eventually  confirmed as the production
of a new heavy resonance, one will presumably be able to measure
its spin and its branching ratios both to  charged  current and to  flavour
and/or lepton number violating channels. 

The possibility of
$s$--channel resonance production  at
HERA  has been suggested in two theoretical schemes. First,
Buchm\"uller and Wyler \cite{BuchWyl} argued that
the  leptoquarks ($\lq$) necessarily  appearing in
grand unified theories  \cite{pati,GG}
can also be accommodated  in theories with conserved lepton and baryon
number. They pointed out that these leptoquarks can be light
(i.e. $m_\lq \ll M_{\rm GUT}$) since low energy
experiments impose much weaker bounds on lepton and 
baryon number  conserving couplings.
Second, the squarks ($\tilde q$) of
 supersymmetric extensions of the Standard Model
 can have direct (lepton number  violating)
 couplings to quarks and leptons if R--parity is not conserved.
Therefore they can also be coupled to lepton--quark
initial states and thus appear as $s$--channel resonances \cite{dreiner1}.
In the present study we shall restrict our discussion
 to  the leptoquark interpretation only,
although several of our results will remain valid also for squark production.

In  section~2 we consider  leptoquark production cross sections at HERA,
including next-to-leading order QCD corrections. From the number of
excess events we extract
the value of the leptoquark couplings ($\lambda$)
and discuss the constraints on these coming from atomic parity
violating experiments.  We also comment  on  non-minimal
models  which allow  for all possible Yukawa
couplings, following the  analysis of ref.~\cite{davidson}.

In section~3 we consider single  and pair production of
scalar and vector leptoquarks in   $p\bar{p}$ collisions at the Tevatron.
 We study  $\lambda $ dependent contributions, 
 the effects of QCD corrections, 
signals and backgrounds, and  experimental limits
on the  leptoquark masses obtained by D0 and CDF.

 Section~4 describes the calculation of the next-to-leading
order QCD corrections to colour-triplet
scalar resonance production in $ep$ collisions.
The numerical value of the  `$K$-factor' and
the average transverse momentum squared of the produced resonance are
also studied.
Full details of the  calculations are given in the  Appendix.
Our conclusions are contained in section~5.

\section{Leptoquark production at HERA and limits on couplings}
\label{sec:two}

We first  consider the minimal scheme in which
the leptoquarks have separate baryon and lepton number conserving chiral
couplings for each family of  mass eigenstates.
This allows us to evade severe
 low energy limits on lepton number violating
couplings \cite{davidson,BRW,miriam}.
Leptoquarks which are coupled to the first family only are called first
generation leptoquarks.
 A leptoquark is said to couple chirally if it couples
either to left-handed (L) or right-handed (R) leptons but not both.

The allowed leptoquark representations have
been classified in  ref.~\cite{BRW}.  There are seven $B$ and $L$ conserving
and SU(3)$\times$SU(2)$\times$U(1)
renormalizable 
 couplings both  for scalar leptoquarks
\bea \label{LQag}
{\cal L}_S&=&  g_{1L}\overline{q}^c_L i \tau_2 l_L S_1
+ g_{1R}\overline{u}^c_R i \tau_2 e_R S_1
+ \tilde{g}_{1R}\overline{d}^c_R i \tau_2 e_R \tilde{S}_1
+ g_{2R}\overline{u}_R  l _L S_2\\[.2cm]\nonumber
&+& g_{2L}\overline{q}_L  e _L S_2
+ \tilde{g}_{2L}\overline{d}_R  l _L \tilde{S}_2
+ g_{3L}\overline{q}^c_L i \tau_2\vec{\tau} l_L S_3 + {\rm h.c.} \ ,
\eea
and vector leptoquarks
\bea
{\cal L}_V&=&  h_{1L}\overline{q}_L \gamma_{\mu} l_L V_1^{\mu}
+ h_{1R}\overline{u}_R \gamma_{\mu} e_R V_1^{\mu}
+ \tilde{h}_{1R}\overline{u}_R \gamma_{\mu} e_R \tilde{V}_1^{\mu}
+ h_{2R}\overline{d}^c_R i\gamma_{\mu} l _L V_2^{\mu}\\[.2cm]\nonumber
&+& h_{2L}\overline{q}^c_L \gamma_{\mu} e _R V_2^{\mu}
+ \tilde{h}_{2L}\overline{u}^c_R \gamma_{\mu} l _L \tilde{V}_2^{\mu}
+ h_{3L}\overline{q}_L \gamma_{\mu}\vec{\tau} l_L V_3^{\mu} + {\rm h.c.} \ .
\eea
We can distinguish the leptoquarks according to their weak isospin
properties:
singlets (e.g. $S_1$), doublets ($S_2$) and triplets ($S_3$).
The doublet scalar leptoquarks and the singlet and triplet vector
leptoquarks
have zero fermion number ($L+3B$) while the singlet and triplet scalar
leptoquark and doublet vector leptoquark have fermion number two. 
Couplings are designated  left or right handed according to the handedness
of the lepton. 

We emphasize  a difference in the theoretical status of scalar and vector leptoquarks.
Within the Standard Model, vector leptoquarks  cannot be considered as gauge
bosons and therefore their couplings to the gauge bosons of the Standard
Model  are not renormalizable in general.
Scalar leptoquarks, however,  
{\it can}  be considered as possible renormalizable matter fields of the
Standard Model. In particular, 
the squarks of the supersymmetric standard model with R--parity
violation provide a
 theoretically appealing  realization of the  isospin doublet
and singlet 
leptoquarks  ($\tilde{S}_2$ and $S_1$).
In contrast,  scalar leptoquarks of  exotic
electric charge ($\tilde{S}_1$), and $S_3$ and vector leptoquarks
 have no counterparts in  supersymmetric models.  
In section~3  it will   be shown that the Tevatron limits on the
production of vector leptoquarks  are only very marginally
consistent with the vector leptoquark
interpretation of the HERA events.
In this section, therefore,  we  shall
consider   limits and production rates  only for  scalar leptoquarks.

The Yukawa couplings in eq.~(\ref{LQag}) are diagonal in the family
number. In order to avoid the severe low energy limits from 
the leptonic decays
of  charged pions   it is convenient to require that 
the leptoquarks   have only one non-zero 
chiral coupling. Thus  we can have left handed or  right handed 
scalar leptoquarks $S_{1R}$, $S_{1L}$ and $S_{2L},S_{2R}$ in (\ref{LQag}).
In Table~1 we list 
 all possible   scalar leptoquark states which can couple
to a $e^+q$ initial state.

The leading order production cross section at HERA has the simple 
form (see Appendix A.1)
\be
\sigma=\frac{\pi\lambda^2}{ 4s } q(m_\lq^2/s) \cdot {\rm BR}\; ,
\ee 
where $\lambda^2$ denotes generically the coupling constants
squared  ($g^2_{L,R}$)
of the produced 
scalar leptoquark as listed in Table~1
(or two times  the coupling constants squared ($h^2_{L,R}$)
of the vector leptoquarks). 
The  cross sections of 
scalar leptoquark production calculated 
to  next-to-leading order (${\cal O}(\alpha_S)$) accuracy are
plotted in Fig.~1 as a function of the leptoquark
mass $m_\lq$ for the $e^+u$, $e^+d$, $e^+s$, $e^+\bar{u}$ and $e^+\bar{d}$    
initial states\footnote{At ${\cal O}(\alpha_S)$ the $e^+g$ scattering
process also contributes, see Appendix, and is included in the cross sections
in Fig.~1.}
 assuming $\lambda^2=1$. For other values of $\lambda$ 
the cross
section obviously scales as $\lambda^2$. The cross section values
vary strongly according to  the initial quark flavour.
For $m_\lq = 200\gev$ we find the ratios (see section~4, Table~3)
\be
\sigma_u:\sigma_d:\sigma_{\bar{d}}:\sigma_{\bar{u}}:\sigma_s=
1440:352:16.7:7.32:8.21
\ee
where the values are given in picobarns.
Assuming 80\% experimental efficiency, we can extract from the H1 and ZEUS 
results  a  rough estimate
for the HERA leptoquark production cross section of $0.7\pb$.
Comparing this value with the predictions of Fig.~1 for $m_\lq=200\gev$
we can calculate numerical values for the various coupling constants
  listed in Table~1, assuming
 branching ratios which correspond to  definite chiral couplings.
\begin{table}
\begin{center}
\begin{tabular}{|l|c|c|c|c|} \hline
process & $\lambda^2(\sigma)$
 & HERA $(g^2)$ & APV $(g^2)$   &   BR  \\ \hline
  $e^+ u \ \to S_2^{5/3}\ \to e^+ u$ & $g_{2R}^2 (+) g_{2L}^2$  &
0.00049&$ < 0.024$ & $1$ \\
  $e^+ d \ \to S_2^{2/3}\ \to e^+ d$ & $g_{2R}^2$ & 0.0020& $<0.024 $ &
    $\frac{g_{2R}^2}{g_{2R}^2 + g_{2L}^2} \to 1$ \\
%
  $\qquad\qquad\qquad  \to \bar\nu u$ &$g^2_{2L}$ &  
   & $-$& $\frac{g_{2L}^2}{g_{2R}^2 + g_{2L}^2} \to 0$ \\
  $e^+ d \ \to \tilde{S}_2^{2/3}\ \to e^+ d$ & $\tilde{g}_{2L}^2$ &
0.0020& $<0.024 $& 
    1 \\
  $e^+ \bar{u}\ \to {S}_1^{1/3}\ \to e^+ \bar{u}$ & 
$g_{1R}^2 (+) g_{1L}^2$ 
   & 0.19, 0.096&$ < 0.012 $ &
 $\frac{g_{1R}^2 + g_{1L}^2}{g_{1R}^2 + 2 g_{1L}^2}\to\frac{1}{2},1$ \\
  $\qquad\qquad\qquad  \to \bar\nu \bar{d}$ & $ g^2_{1L} $   &  & $-$ &
    $\frac{g_{1L}^2}{g_{1R}^2 + 2 g_{1L}^2}\to \frac{1}{2},0 $\\
  $e^+ \bar{d} \ \to \tilde{S}_1^{4/3}\ \to e^+ \bar{d}$ & $\tilde{g}_{1R}^2$ 
   &0.0420& $<0.024$ & 1 \\
  $e^+ \bar{u}\ \to {S}_3^{1/3}\ \to e^+ \bar{u}$ & $g_{3L}^2$ 
   &0.096&$<0.024$&  $\frac{1}{2}$ \\

  $\qquad\qquad\qquad  \to \bar\nu \bar{d}$ &  $g_{3L}^2$ 
   &  & & $\frac{1}{2}$ \\
  $e^+ \bar{d} \ \to {S}_3^{4/3}\ \to e^+ \bar{d}$ & 2 $g_{3L}^2$ 
   &0.021&$-$ & 1 \\
\hline
\end{tabular}
\caption{Scenarios for scalar leptoquark production in $e^+p$ collisions,
with values and limits for the various couplings obtained from HERA and
atomic parity violation (APV). Limiting values for the branching ratios
corresponding to definite chiral couplings (i.e. L or R) are given.} 
\end{center}
\end{table}

In the the minimal scheme  the strict bounds on
electron and muon number conservation are automatically satisfied.
Allowing only one chiral coupling, 
 the severe constraints from  the leptonic decays of charged pions and kaons
are also avoided.
But in the case of first generation leptoquarks
the diagonal flavour conserving couplings (see  Table~1) are also
severely limited by the precise measurements of the atomic parity
violating
neutral current weak charge of the caesium atom
\be
Q_W=-376C_{1u}-422 C_{1d} \, ,
\ee
where $C_{1u}$ and $C_{1d}$ denote the
lepton axial vector --  quark vector interference terms of the effective 
low energy four fermion interaction
\be
{\cal L}_{\rm APV} = 
\frac{G}{\sqrt{2}}\sum_{i=u,d} C_{1i}\, \overline{l}\gamma^{\mu}\gamma^5 l
\, \overline{q}_i\gamma_{\mu}q_i \, ,
\ee 
with
\be
C_{1u}=-\frac{1}{2}+\frac{4}{3}\sin^2 \Theta_W\, ,\qquad
C_{1d}=\frac{1}{2}-\frac{2}{3}\sin^2 \Theta_W\, .
\ee
The scalar leptoquark contribution to $C_{1q}$ is
\be
\Delta C_{1q}=\frac{\sqrt{2}g^2}{8G_F m^2_{\lq}} \, .
\ee
The measured value is  \cite{rpp96}
\be
Q_W=-71.04\, \pm\, 1.58\, \pm\, [0.88]_{\rm th.}
\ee
to be compared with the Standard Model prediction \cite{rpp96}
\be
Q_W=-72.88\,\pm\, 0.05\,\pm\, 0.03\, .
\ee
The difference between  theory and experiment is therefore smaller than
\be
\abs{\Delta Q_W} = 3.6 \, ,
\ee
which gives 
\be
g^2<0.024 \left(\frac{m_{\lq}}{200\gev}\right)^2
\ee
which in turn leads to the values given in the fourth column of Table~1.
From this we conclude that the HERA and the APV data are  consistent
{\it only with the production of isodoublet (zero fermion number)
leptoquarks}. The parton distributions for antiquarks are too small
at large $x$ ($\sim 1/2$) 
to allow sizeable production of leptoquarks with the weak coupling required by 
 atomic parity violation data.
If this interpretation is correct and only one leptoquark
exists in the HERA kinematic range, then it follows that
in $e^-p$ collisions the signal will be  weaker by about two orders of
magnitude!
While this is an attractive explanation by itself, it may be excluded
soon by   new analyses from CDF and D0 at the Tevatron. We shall see
in the
next section that 
 scalar leptoquarks of mass    $m_\lq \sim 200\gev$ are 
very close to the published discovery limit for leptoquark pair
production at the Tevatron.

The Tevatron limit can be weakened by suppressing
the  branching ratio into charged lepton $+$ jet,
since the production is essentially independent of the leptoquark coupling.
This motivates us to consider the more general parametrization
of leptoquark production as studied by Davidson \etal\ \cite{davidson},
 with general mixed couplings between
the three families.
In the Lagrangian  ${\cal L}_{S}$ of  (\ref{LQag}) the family diagonal
leptoquark couplings $g$
 are replaced by the coupling matrix $g^{ij}$ where the summation labels run
over the family indices $i,j=1,2,3$.
Davidson \etal\  \cite{davidson}  summarize their
results\footnote{See Table~15 of ref.~\cite{davidson}.}
in terms of bounds on coupling constant combinations
$$\frac{1}{2}g^{ij}g^{km}\left(\frac{m_{\lq}}{100\gev}\right)^2$$
In Table~2 we reproduce their results
for $g^{1j}g^{mn}$ at $m_{\lq}=200\gev$.

\begin{table}
\begin{center}
\begin{tabular}{|c|c|c|c|c|c|c|c|} \hline
$g^{1j}g^{mn}$ & $S_{1L}$ &$ S_{1R}$ &$\tilde{S}_{1R}$&
 $S_{2L}$  &  $S_{2R}$& $\tilde{S}_{2L}$&$ S_{3L}$   \\ \hline
(11)(11) &.006 & .008 & .008 
& .008 & .004 & .008 & .004 \\
(11)(12)& 
$4\times 10^{-5}$&
 .012 
&$ 6\times 10^{-4}$
& $0.012$ 
&$6\times 10^{-4}$
&$4\times 10^{-5}$
&$4\times 10^{-5}$ \\
(11)(13)&.008 &*&.012& *& .012& .012 & .008\\
(11)(21)
&$2\times 10^{-6}$&$2\times 10^{-6}$&$2\times 10^{-6}$
&$2\times 10^{-6}$&$2\times 10^{-6}$&$2\times 10^{-6}$
&$6\times 10^{-7}$\\
(11)(23) &.008 &* &.012 &* &.012 &.012 &.008\\
(11)(31) & .006&$2\times 10^{-2}$
 &$2\times 10^{-2}$ &$2\times 10^{-2}$ &$2\times 10^{-2}$ &
$2\times 10^{-2}$&.006\\
(11)(32) &$4\times 10^{-5}$ & &.1 & &.1 &
$4\times 10^{-5}$ &$4\times 10^{-5}$\\
(11)(33) &.008 & * &.16 &* &.16&.16 &.008\\
(12)(12) &1.2 & & & & & &1.2\\
(12)(13) &.08 & &$1.2\times 10^{-3}$& &$1.2\times 10^{-3}$ &
$1.2\times 10^{-3}$ & $6\times 10^{-4}$\\
(12)(21) &$4\times 10^{-5}$ &.2 &$2\times 10^{-6}$ &.2 &
$2\times 10^{-6}$ &$2\times 10^{-6}$ &$1.2\times 10^{-6}$\\
(12)(22) &$8\times 10^{-5}$ &$8\times 10^{-5}$ &
$8\times 10^{-5}$ &$8\times 10^{-5}$ &$8\times 10^{-5}$ &$8\times 10^{-5}$ &$8\times 10^{-5}$\\
(12)(23) &.08 &* &$1.2\times 10^{-2}$ &* &$1.2\times 10^{-2}$ &
$1.2\times 10^{-2}$ &$6\times 10^{-3}$\\
(12)(31) &$4\times 10^{-5}$ & & .1& &.1 &$4\times 10^{-5}$ &$4\times 10^{-5}$\\
(12)(32) &1. &1. &.6 &.4 &.4 & &1.\\
(12)(33) &.08 &* &.16 &* &.16 &.16 &.08\\
\hline
\end{tabular}
\caption{Generation--mixing leptoquark coupling limits, 
from ref.~\protect\cite{davidson}. The interactions
denoted by $*$ involve a top quark and the corresponding constraints have to
be evaluated using the known top quark mass.}
\end{center}
\end{table}
We see from  Table~2 that there are many  ways
to avoid the bounds and still  maintain consistency with the HERA signal.
It
is still premature to make a detailed study, therefore 
we mention only  one possibility.
Suppose that all the couplings are extremely small except
(12)(12), i.e.  when  the leptoquark is coupled  to a strange quark
and a positron or to a charm quark and a positron.
The smaller strange quark and charm quark content of the
proton requires larger leptoquark couplings (see Fig.~1) but atomic
parity violation constraints are avoided since there is no coupling to
a second generation quark in the proton.
In this scenario the signal in $e^-p$ scattering  will be as large
as for  $e^+p$. If we insist on a   smaller  branching
ratio ($<1$) then
we may allow either for both left and right handed couplings or
for a  significant (12)(23) coupling. Another possibility
is to allow also for a (12)(31) coupling. In this way the Tevatron
mass limit could be relaxed. In this scenario charged current decay
modes can also easily be arranged.
Finally we note  that (i) the (12)(12) solution resembles the squark
solution suggested in  \cite{xxx}, and (ii) 
although the above  choice may appear somewhat {\it ad hoc}  we 
should recall the huge hierarchy of the Yukawa couplings  
in the Standard Model without scalar leptoquarks.

\section{Leptoquark production at the Tevatron}

Colour--singlet leptoquarks have a standard gauge coupling ($g_s$)
to gluons and can therefore be copiously pair produced in hadron--hadron
collisions. The direct coupling to quarks ($\lambda$)
gives additional contributions to the production cross section, but these
are numerically much less important for the range of $\lambda$ allowed
by the HERA and APV data. In addition, {\it vector} leptoquarks can have
additional anomalous couplings  ($\kappa_V,\lambda_V$)  to gluons, analogous
to the anomalous electric quadrupole and magnetic dipole
 moments of $W$ bosons. Although
these lead to a violation of unitarity in the production cross section,
they may be regarded as an effective low energy parametrization of a more
complicated theoretical structure.

Assuming that $\lq \to l + q$ is the dominant leptoquark decay mode,
the signature of pair production in hadron collisions -- two energetic 
leptons and two energetic jets widely separated in phase space -- 
is rather distinctive. Backgrounds from processes like $W,Z+$jets,
$b \bar b+$jets, etc.
can be suppressed in principle by kinematic cuts.

The pair production 
subprocess cross sections are\footnote{The cross sections for 
scalar pair production have been derived in the literature in many
different guises. See, for example, ref.~\cite{Blumlein96} and
references therein.},
 with $\beta^2 = 1-4m_{\lq}^2/\hat{s}$,
\begin{eqnarray}
\hat{\sigma}_{gg\to S \bar S} &=& { \pi \alpha_S^2 \over 96 \hat{s}}
\; \left[ \beta (41-31\beta^2) -(17-18\beta^2 + \beta^4) 
\ln\left( { 1+\beta \over 1 - \beta } \right) \right] 
\; , \nonumber \\
\hat{\sigma}_{q\bar q\to S \bar S} &=& {2 \pi \alpha_S^2 \over 27 \hat{s}}
\; \beta^3 \; , 
\end{eqnarray}
for scalar pair production and
\cite{Blumlein96,Arnold86,Borisov87,Blumlein93,HRPHP}
\begin{eqnarray}
\hat{\sigma}_{gg\to V \bar V} &=& { \pi \alpha_S^2 \over 96 m_\lq^2}
\;  \Biggl[
\beta \left(\frac{523}{4} -90\beta^2 + \frac{93}{4}\beta^4\right)
 \nonumber  \\& &  \qquad \qquad
-\frac{3}{4}\left(65-83\beta^2 + 19\beta^4-\beta^6\right) 
\ln\left( { 1+\beta \over 1 - \beta } \right) \Biggr]  \; , \nonumber \\
\hat{\sigma}_{q\bar q\to V \bar V} &=& { 4\pi \alpha_S^2 \over 9 m_\lq^2}
\; \frac{\beta^3}{24} \left[ 23 - 3 \beta^2 + 
\frac{4}{1-\beta^2} \right]   \; , 
\label{vvpair}
\end{eqnarray}
for vector pair production, with $\lambda_V = \kappa_V = 0$. 
Note that in general 
the scattering amplitudes for $q \bar q \to S\bar{S},V\bar{V}$ receive
additional contributions ($\hat{\sigma} \sim \alpha_S \lambda^2, \lambda^4$)
from diagrams involving direct quark--leptoquark couplings (e.g.
$q \bar q \to \lq\overline{\lq}$ by $t$--channel exchange of a lepton).
However for $\lambda^2 = {\cal O}(10^{-1})$ or smaller, these contributions
are much smaller than the ${\cal O}(\alpha_S^2)$ contributions listed above.
The generalization of (\ref{vvpair}) including non-zero anomalous
couplings is presented in ref.~\cite{Blumlein96}.

Fig.~\ref{fig:tev} shows both cross sections
at $\sqrt{s} = 1.8$~TeV as a function of the leptoquark mass. Note that
for $m_{\lq} \sim 200\gev$ the $q\bar q$ annihilation subprocesses are dominant.
The parton distributions are the MRS(R2) set from ref.~\cite{mrsr}
(with $\alpha_S(M_Z^2) = 0.12$) and the renormalization and factorization scales
are set equal to $m_{\lq}$. We see from  Fig.~\ref{fig:tev} that the 
cross sections for $m_\lq \sim 200\gev$ vector and scalar pair production are
 ${\cal O}(10)\pb$ and  ${\cal O}(0.1)\pb$ respectively. The former are therefore
 ruled out by the current Tevatron data \cite{D0leptoq}.
 The only caveat to this is that it is possible to `fine-tune' the anomalous
 couplings $\kappa_V,\lambda_V$ to suppress the vector pair cross section
 by between one and two orders of magnitude, see
for example ref.~\cite{Blumlein96}.
However this would appear to be completely unnatural. The cross sections
in Fig.~\ref{fig:tev} do not include NLO QCD corrections. Estimates for these
can be extracted from the calculation of ref.~\cite{kfac} for squark
pair production in the infinite gluino mass limit. Numerically, the NLO
corrections increase the cross section by a modest ${\cal O}(+10\%)$ in the mass range
of interest (see Fig.~19(b) of ref.~\cite{kfac}).
  
A $200\gev$ scalar leptoquark would therefore give rise to approximately
10 events in a sample corresponding to $100\pb^{-1}$ at the Tevatron.
In fact a recent  detailed analysis by the D0 collaboration \cite{D0leptoq} of 
their combined Run~I data gives a ($95\%$cl) lower mass limit of
$175\gev$ ($147\gev$)
assuming ${\rm BR}(\lq\to e q) = 1 (0.5)$.\footnote{There is as yet
no combined Run~I published first generation leptoquark mass
limit from the CDF collaboration.}
 Thus the
$m_\lq \sim 200\gev$ (scalar $\lq$) interpretation of the HERA excess events
is allowed, by a small margin. Obviously the Tevatron limit is weakened
if the branching ratio to $eq$ is lowered (see section~2 above).

Finally, we show also in Fig.~\ref{fig:tev}
{\it single} scalar leptoquark production cross sections
from the subprocess $qg \to \lq + l$ \cite{Hewett88}:
\begin{equation}
\hat{\sigma}_{qg} = {\lambda^2 \alpha_S\over 48 \hat{s}}\;
\left[ 1 + 6x - 7 x^2 + 4x(1+x)\ln x\right] \; ,
\end{equation}
with $x=m_{\lq}^2 / \hat{s} \leq 1$.
In Fig.~\ref{fig:tev} we have taken, for illustration,
$\lambda^2 = 0.01$ and distinguished the cases
where the leptoquark couples to $u$, $d$ and $s$ quarks.
Thus, for example, for a scalar leptoquark which is
produced at HERA via $e^+u$ and has a 100\% branching ratio
into the same state, the curve labeled `$S_u$' gives the number
of  $g u \to (e^+u) + e^-$ and
$g \bar u \to (e^- \bar u) + e^+$ events. Other $\lambda^2$ values
and branching ratios can be obtained by a simple rescaling of the
curves. Note also that for a scalar leptoquark which gives
rise also to charged current events are HERA, for example
via $e^+ d \to S \to \bar \nu_e u$, the additional `missing energy'
processes $ g d \to (\bar \nu_e u) + e^-$ and $ g u \to (\bar \nu_e u) +
\nu_e$ are possible. The corresponding event rates are readily
estimated from the curves in Fig.~\ref{fig:tev}.

The most important conclusion from the single leptoquark production
curves in Fig.~\ref{fig:tev} is that for $m_{\lq} \sim 200\gev$,
the single production cross section is less than the pair production
cross section for $\lambda^2 < {\cal O}(0.07, 0.16, 0.8)$ for coupling
to $u,d,s$ quarks.

\section{Production cross section in NLO at HERA}

In order to get an accurate value of the leptoquark
coupling constant from the HERA  data one should take into account
 the next-to-leading order QCD 
corrections. This is also required  for consistency if
parton distributions
obtained with a NLO fitting procedure are used.
 In this section we describe the calculation for {\it scalar}
leptoquarks since, as we have seen, vector particles are disfavoured
by searches at the Tevatron.
Note that the transverse momentum ($k_T$) distribution of the leptoquarks
with respect to the beam direction is a by-product of the full NLO correction.

The NLO-corrected cross section can be written as 
\begin{eqnarray}
\sigma(e^+q) &=& \frac{\pi\lambda^2 }{4 s}  \int_x^1 \frac{dz}{z}
\left[ q\left(\frac{x}{z},\mu_{IR}^2\right) \left\{ \delta(1-z) + 
{\alpha_S(\mu^2)\over 2 \pi}
C_F K_q(z)\right\} \right. \nonumber \\
& & \qquad \qquad\left. 
+ g\left(\frac{x}{z},\mu_{IR}^2\right){\alpha_S(\mu^2)\over 2 \pi}
T_R K_g(z)\right] \; .
\label{signlo}
\end{eqnarray}
where $x = m_\lq^2/s$ and  $C_F = 4/3 $,  $T_R = 1/2$.
The coefficient functions are calculated from the ${\cal O}(\alpha_S)$
Feynman diagrams. Using $\overline{\rm MS}$ factorization and
renormalization we obtain (see Appendix for more details)
\begin{eqnarray}
K_q(z) & = & \delta(1-z) \left( - {\pi^2\over 3}
+ \frac{3}{2} \ln{\mu_{UV}^2 \over \mu_{IR}^2} \right)- {2z\over (1-z)_+}
\; ,  \nonumber \\
& & \qquad 
      + 2(1+z^2) \left({\ln(1-z)\over 1-z} \right)_+ 
-\ln\left({z\mu_{IR}^2\over m^2}\right) {1 + z^2 \over (1-z)_+ }  \\
K_g(z) & = &  \left[ (1-z)^2+z^2\right] \ln\left(
{z(1-z)^2\mu_{IR}^2\over m^2}\right)
+2z(1-z)(2+\ln z)  \; .
\end{eqnarray}
 Here $\mu_{IR}^2$ and $\mu_{UV}^2$ denote
the factorization and (ultraviolet) renormalization\footnote{The leptoquark
coupling $\lambda^2$ is renormalized by the ${\cal O}(\alpha_S)$
vertex corrections, see Appendix, 
hence $\lambda^2 \equiv \lambda^2(\mu_{UV}^2)$.} scales respectively.
In our numerical calculations we will use
\begin{equation}
\mu^2 = \mu_{IR}^2 = \mu_{UV}^2 = m_\lq^2\; .
\end{equation}
Note that the QCD correction is universal for the different types of scalar
leptoquarks discussed in section~1. 

\begin{figure}[p]
\begin{center}
\mbox{\epsfig{file=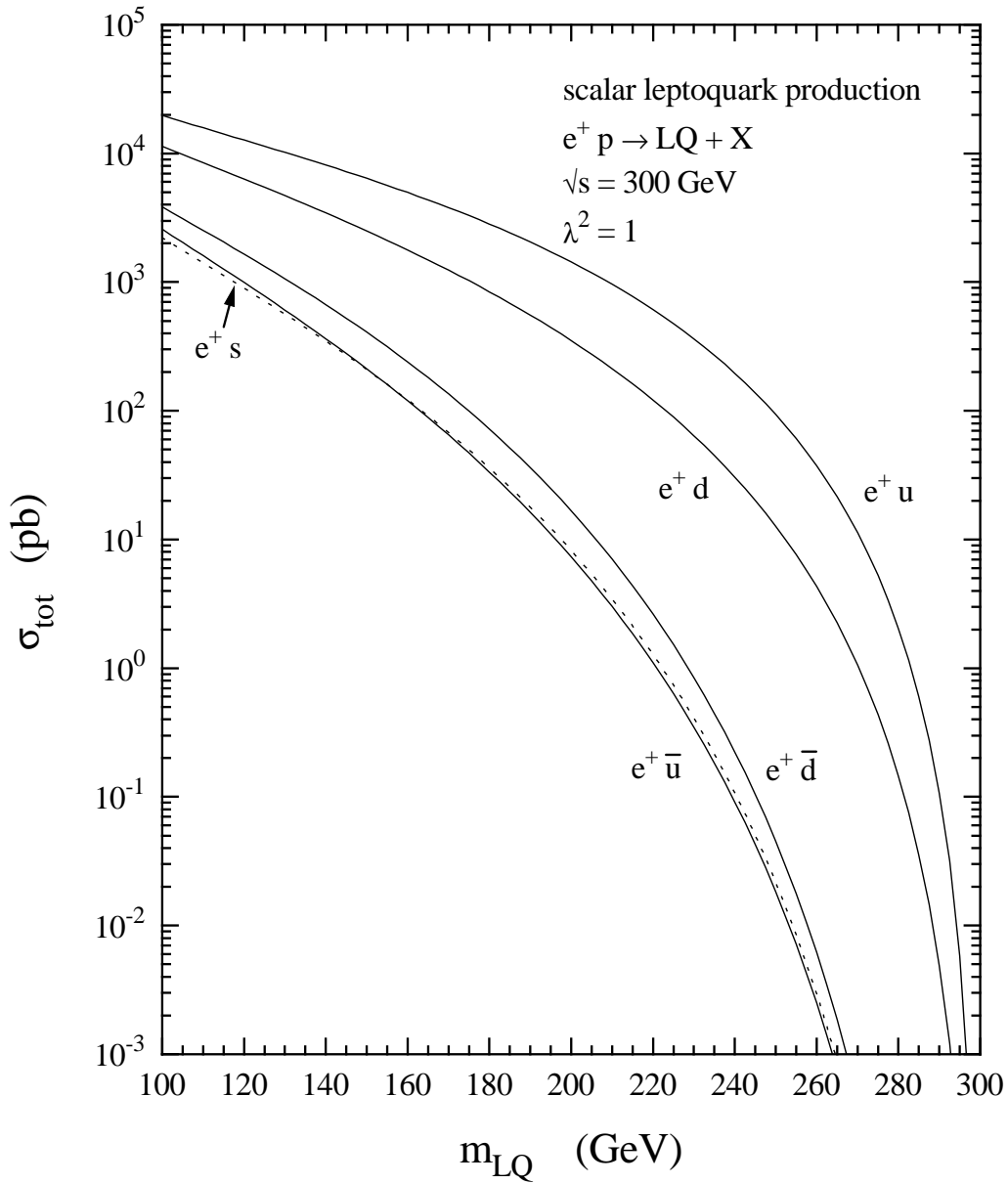,height=20cm}}
\caption{Scalar leptoquark production cross sections in NLO QCD
for $e^+p$ collisions at $\protect\sqrt{s} = 300\gev$.}
\label{fig:sig}
\end{center}
\end{figure}

Figure~\ref{fig:sig}
 shows the NLO--corrected production cross sections for
$e^+ q \to \lq$ with $q = u, d, \bar u, \bar d, s$ at $\sqrt{s} =
300\gev$.
For ease of comparison we have set the overall leptoquark coupling
to unity, i.e.  $\lambda^2 = 1$. The parton distributions 
are the MRS(R2) set from ref.~\cite{mrsr},
with $\alpha_S(M_Z^2) = 0.120$. 
The cross section hierarchy  mainly reflects
the hierarchy of the quark  distributions at large $x$.
 The ratio of the NLO to LO
contributions (the `K--factor' ${\cal K}$) is non-negligible. For the valence
quark cross sections, ${\cal K}$ increases from about $1.25$ to $1.45$ in the
 mass range of interest, $175\gev < m_\lq < 225\gev$, while for the
 sea quarks ${\cal K}$ increases from about $1.3$ to $1.6$ in the 
 same mass range. Table~\ref{tabsigma} lists  cross section values
 for leptoquark masses relevant to the HERA high $Q^2$ events.
In the limit $m_\lq \to \sqrt{s}$, i.e. $x \to 1$, the correction
 is dominated by a soft-gluon double logarithm, 
${\cal K} \sim  1+ \alpha_S C_F \ln^2(1-x)/\pi$.

\begin{table}[htb]
\begin{center}
\begin{tabular}{|l|c|c|c|c|c|} \hline
 & $\sigma(e^+u)$  & $\sigma(e^+d)$ 
     & $\sigma(e^+\bar u)$  & $\sigma(e^+\bar d)$  & $\sigma(e^+ s)$ \\ \hline
$m_\lq = 175\gev$ & 3289   &   1026  &  47.0    &  99.6    &   49.7  \\
              & (2659)   &   (801)  & (35.7) &  (73.8) &  (37.9) \\ \hline    
$m_\lq = 200\gev$ & 1436  &    352   &   7.32 & 16.7   &  8.21     \\
             & (1096) &   (260)  &  (5.20) & (11.5) &  (5.82)        \\ \hline
$m_\lq = 225\gev$ & 475   &   89.1   &   0.634 & 1.52    &  0.740      \\
              & (337) &  (61.3)  &   (0.413)  &  (0.96) &   (0.479)   \\ \hline
\end{tabular}
\caption{Leptoquark cross sections (in pb) at NLO in $e^+p$ collisions
at $\protect\sqrt{s} = 300\gev$,
assuming unit coupling $\protect\lambda^2 = 1$.
The numbers in brackets are the corresponding leading-order
cross sections.}
\end{center}
\label{tabsigma}
\end{table}

\begin{figure}[p]
\begin{center}
\mbox{\epsfig{file=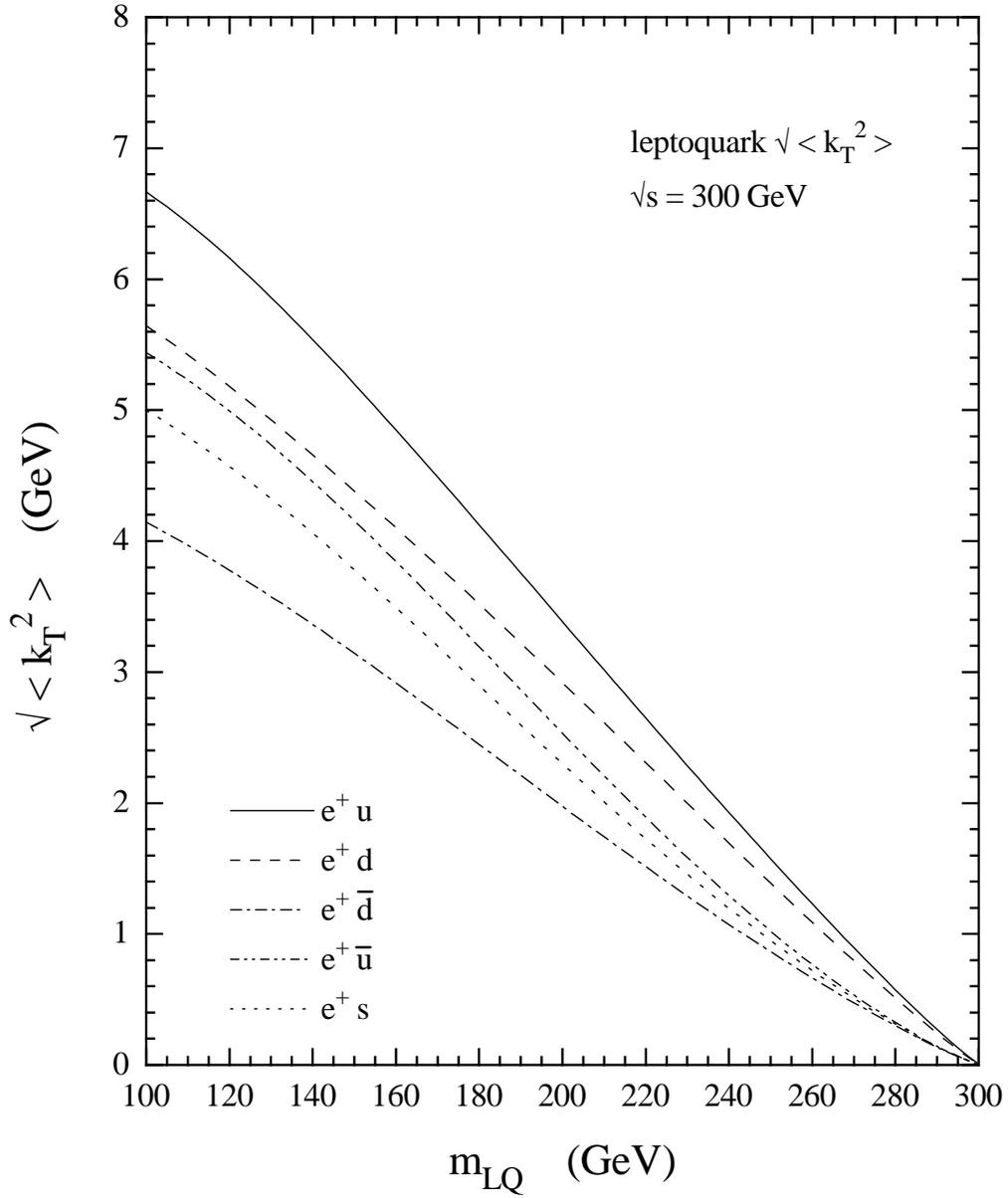,height=20cm}}
\caption{Average scalar leptoquark transverse momentum
 for $e^+p$ collisions at $\protect\sqrt{s} = 300\gev$.}
\label{fig:kt}
\end{center}
\end{figure}

\begin{figure}[p]
\begin{center}
\mbox{\epsfig{file=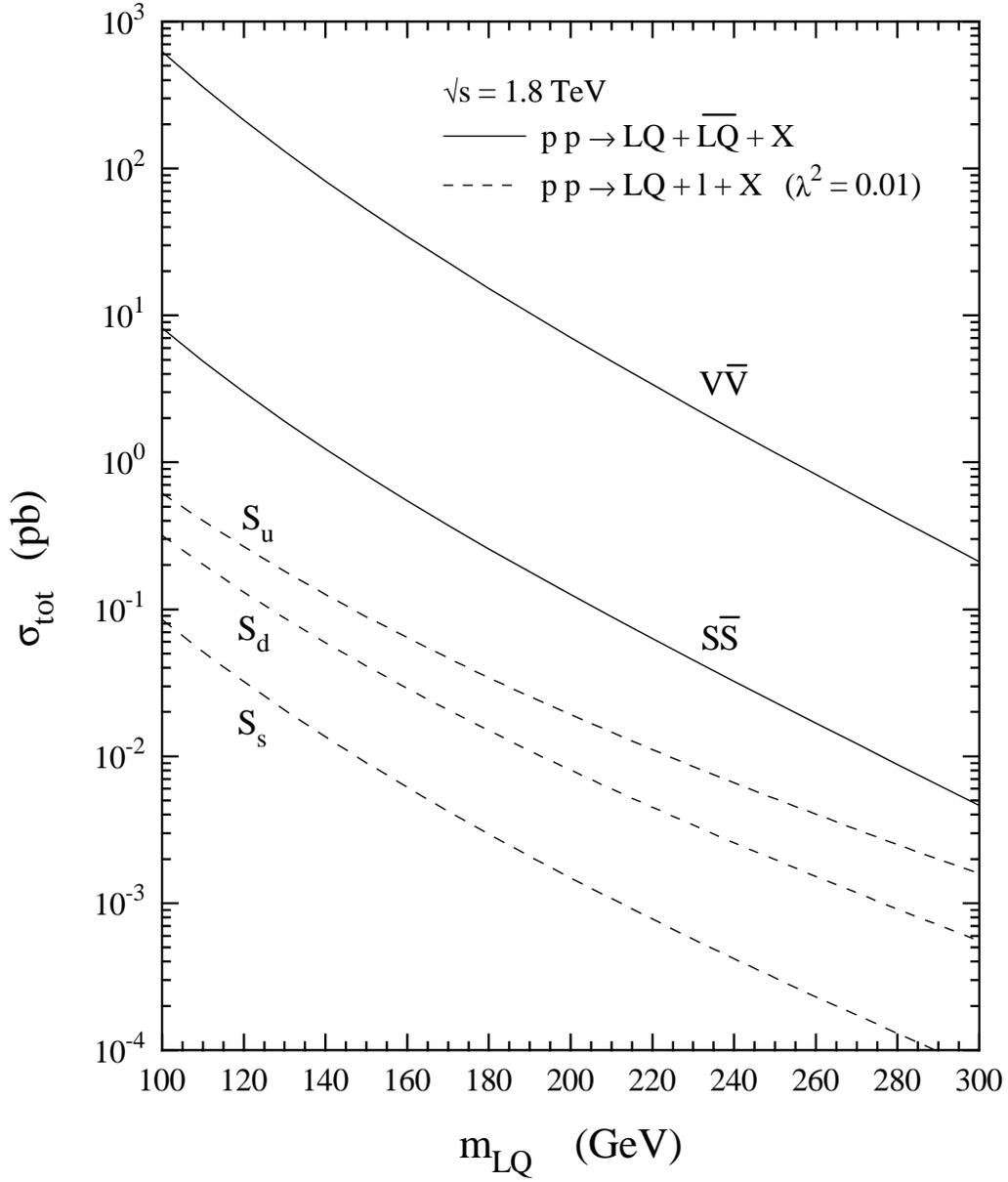,height=20cm}}
\caption{Scalar and vector leptoquark production at the Tevatron
$p \bar p$ collider.}
\label{fig:tev}
\end{center}
\end{figure}

The transverse momentum distribution $d\sigma/dk_T^2$ and the 
average transverse momentum squared $\langle k_T^2 \rangle$ of the leptoquarks
can also be calculated from the $O(\alpha_S)$ diagrams 
for the processes $eq \to g \lq $ and $eg \to q \lq  $ (see Appendix~A.4).
For the average we obtain
\begin{equation}
\label{eqkt}
{\langle k_T^2 \rangle \over m_\lq^2} =
\frac{\pi\lambda^2 }{4 s \sigma}\; {\alpha_S(\mu^2)\over 2 \pi} \;
 \int_x^1 \frac{dz}{z^2}
\left[ q\left(\frac{x}{z},\mu_{IR}^2\right) C_F \overline{K}_q(z)
+ g\left(\frac{x}{z},\mu_{IR}^2\right)T_R \overline{K}_g(z) \right] \; ,
\end{equation}
where $\sigma$ is defined in eq.~(\ref{signlo}) and 
\begin{eqnarray}
\label{eqktq}
\overline{K}_q(z) & = &  \frac{1}{3}\; (1+z^2)(1-z) \; , \\
\label{eqktg}
\overline{K}_g(z) & = &   \frac{1}{2}
-3z+\frac{z^2}{2} + 2 z^3 - 4 z^2 \ln z  \; .
\end{eqnarray}
Figure~\ref{fig:kt} shows $\sqrt{ \langle k_T^2 \rangle }$ as a function
of $m_\lq$ for the various $e^+q$ production mechanisms. Note that the average
transverse momentum is in general very small, because the emission
of an additional energetic parton is heavily suppressed.

\section{Conclusions}

We have calculated the NLO QCD corrections to the  production
cross-section of a
coloured s-channel scalar resonance (leptoquark or squark)  
at HERA and  reevaluated various leptoquark production cross sections for
Tevatron. The average transverse momentum of the leptoquark produced
at HERA was also calculated from the NLO result.
Our calculation was motivated  by the
leptoquark interpretation of the excess of events found by H1 and ZEUS
at HERA in $e^+p$ deep-inelastic scattering
above the Standard Model prediction.
The simplest and perhaps most attractive `new physics'
interpretation of the data is in terms of a first family leptoquark with
a single Yukawa coupling.
 Constraints from data on atomic parity
violation, double beta decay and various flavour changing processes
are consistent with the HERA data only for
 isodoublet scalar leptoquark production. 
This interpretation, however, is seriously  challenged
by   severe constraints from leptoquark pair
production
at  the
Tevatron $p \bar p$ collider
 which is largely independent from the Yukawa coupling
of the leptoquark. The Tevatron limits are even more
severe for vector leptoquarks. Because the QCD corrections to
the HERA cross sections are large and positive, the extracted
couplings are somewhat smaller than those obtained from a leading
order analysis only.

If one also considers leptoquarks which couple to several
generations, then in the enlarged  space of couplings one can find    several
solutions  compatible with  the  low energy  constraints  and the HERA
excess events.
In particular, one can have solutions with 
reduced branching ratios and/or
charged current events, as well as lepton
number violating processes.  Furthermore  one can no longer rule out
the isosinglet and isotriplet scalar leptoquark interpretation in such
models.
The solutions with reduced branching ratios 
 also relax the Tevatron limits.
However, they do require
a rather  {\it ad hoc} hierarchy in the Yukawa couplings
(although such a strong hierarchy already exists in the
Standard Model Higgs couplings).

The strong hierarchy in the Yukawa couplings in the allowed solutions
implies a  very specific flavour
structure and therefore such models can be tested
when more HERA data are available.
We have considered, as a specific example, the case
where the leptoquark is dominantly
coupled to a positron and a strange squark.

\vskip 1truecm

\noindent {\it Note added:} Since the H1 and ZEUS results were made public,
there have been many theoretical papers attempting to explain the
apparent event excess in terms of leptoquark--type particles \cite{many}.
The present study overlaps with some of these in some respects, but to our 
knowledge our calculation of the HERA cross sections is the first to 
include the ${\cal O}(\alpha_S)$ QCD corrections.

\newpage
\section*{Appendix}
\setcounter{equation}{0}
\setcounter{section}{1}
\renewcommand{\theequation}{\Alph{section}\arabic{equation}}
\renewcommand{\thesection}{\Alph{section}}

\subsection{Kinematics and leading-order cross section}
\label{sec:kinematics}

We consider the process 
\be\label{process}
e^+(p_2)\ + \ q(p_1)\  \to \  R(q)
\ee
\be
p_1\ + p_2\ = \ q\,,\qquad \hat{s}=(p_1+p_2)^2 = q^2 = m^2\,,\qquad
 p_1^2 = p_2^2 = 0 \,.
\ee
The leading order transition amplitude is
\be\label{TB}
T_{\rm B} =\lambda \overline{v}(p_2)\,P_{\rm L,R}\,u(p_1)\,,
\quad P_{\rm L,R}=\frac{1 \pm \gamma_5}{2}\,.
\ee
The spin- and colour-averaged squared amplitude is
\be
\overline{\abs{T_{\rm B}}}^2 = \frac{\lambda^2 \hat{s}}{4}\,,
\ee
which  gives the leading-order cross sections
\be\label{sigma0}
\hat{\sigma}_{0}(\hat{s})=\frac{\pi\lambda^2}{4}\delta(\hat{s}-m^2)
\ee
\be
\sigma_{\rm ep}=\int dy\,  q(y)\, \hat{\sigma}_0(\hat{s} = ys)=
\frac{\pi\lambda^2}{ 4s } q(x)\,, \quad \mbox{with}\ \  m^2=x s\,.
\ee 
\subsection{Virtual corrections}
\label{sec:VC}
In the calculation of the real and virtual corrections, we regulate
the infra-red and ultraviolet divergences using dimensional
regularization with $d=4-2\ep$.
\subsubsection{Loop integrals}
\noindent
We need the loop integrals over two propagators:
\bea\label{loopintI2}
I_2(q^2)&=& \int d^dk\frac{1}{(k^2+i\eta)\left[(k-q)^2+i\eta\right]}\, =\, 
 Q_{-q^2}T_{0}^{\rm UV}\,,\ \ \  q^2\ne 0\,,\\[.3cm]
&&Q_{-q^2}=i\pi^{d/2}\Gamma(1+\ep)(-q^2-i\eta)^{-\ep}\,,\\[.2cm]
&&
T_0^{\rm UV,IR}=\frac{1}{\ep_{\rm UV,IR}} + 2\,,\\*[.2cm]
I_2(q^2)&=&i\pi^2\frac{(4\pi)^{\ep}}{\Gamma(1-\ep)} 
\left(T_0^{\rm UV}\,-\,T_0^{\rm IR}\right)
 = 0 \quad {\rm if}\quad q^2=0\,,\\*[0.2cm]
I_2(q)_{\mu}&=& \int d^dk
\frac{k_{\mu}}{(k^2+i\eta)\left[(k-q)^2+i\eta\right]}\, =\,
 \,\frac{1}{2} q_{\mu} I( q^2 )\,,\\*[0.2cm]
I_2(q^2;m^2)&=& 
\int d^dk\frac{1}{(k^2+i\eta)\left[(k-q)^2-m^2+i\eta\right]}
\\*[0.2cm]\nonumber& =&
 Q_{m^2}\frac{1}{\ep}\int_0^1 dx 
\left[x-x(1-x)\frac{q^2}{m^2}\right]^{-\ep}\,,
\\*[.2cm]
I_2(m^2;m^2)&=&Q_{m^2}T_{0}^{\rm UV}\,,\\*[.2cm]
I_2^{\,'}(m^2;m^2)&=&
m^2\frac{dI_2(q^2;m^2)}{dq^2}{\Big\vert}_{q^2=m^2}\,=\,-Q_{m^2}
\frac{1}{2}T_{0}^{\rm IR}\,,
\eea 
and the loop integral over three propagators:
\bea\label{loopintI3}
I_3(q,p)&=& \int d^dk\frac{1}{(k^2+i\eta)\left[(k-q)^2+i\eta\right]
\left[(k-p)^2+i\eta\right]}\,,\\*[.2cm]\nonumber &&
\quad  q^2\ne 0\,,p^2=0\,,(q-p)^2=0\,,\\*[.2cm]
I_3(q,p)&=&-Q_{-q^2}\left(-\frac{1}{q^2}\right)\, R_0\,,\quad 
R_0\,=\,\frac{1}{\ep^2}\,-\,\frac{\pi^2}{6}+ {\cal O}(\ep)\,,\\*[.2cm]
I_3(q^2;m^2)&=& \int d^dk\frac{1}{(k^2+i\eta)\left[(k-q)^2-m^2+i\eta\right]
\left[(k-p)^2+i\eta\right]}\,,\\*[.2cm]\nonumber &&
\quad  q^2\ne 0\,,p^2=0\,,(q-p)^2=0\,,\\*[.2cm]
I_3(m^2;m^2)&=&\frac{1}{m^2}Q_{m^2}\left(-\frac{1}{2\ep^2}\right)\,.
\eea
It is interesting  to compare the massless and massive $I_3$
integrals:
the  double pole singularities have opposite sign and different
normalization,
and the finite terms are also different.
We note also  the relation
\be\label{gam+gam-}
\Gamma(1+\ep)=
\frac{1}{\Gamma(1-\ep)}
\left( 1  +\ep^2
\frac{\pi^2}{6}\right)
 + {\cal O}(\ep^3)\,.
\ee
\subsubsection{Self-energy corrections}
\noindent
Self-energy corrections give wave-function renormalization through the
relations
\be
T_{\rm S}=\frac{1}{2}T_{\rm B}
\frac{d\Sigma_{\rm S}}{dq^2}{\Bigg\vert}_{q^2=m^2}\,,
\qquad 
T_{\rm F}=
\frac{1}{2}T_{\rm B}\frac{d\Sigma_{\rm F}}{d\rlap{p}/}{\Bigg\vert}_{p^2=0}
\ee
where $T_{\rm B}$ denotes the Born amplitude (see eq.~(\ref{TB})) and
$-i\Sigma_{\rm S}$, $-i\Sigma_{\rm F}$ are given by the self-energy
diagrams for massive scalar and massless fermion lines respectively.
Introducing the auxiliary notation
\be
D_1=(k-q)^2-m^2\,,\quad D_2=(k-p_1)^2\,,\quad D_3=k^2\,,\quad
C=-iC_F(2\pi)^{-d}g_s^2\mu^{2\ep}\,,
\ee
where $C_F=4/3$ is the usual colour factor, we can write
\bea 
\Sigma_{\rm S}&=&C\int d^dk\frac{(2q-k)^2}{D_1 D_3}=
C\int d^dk\frac{2D_1-D_3+2(q^2+m^2)}{D_1 D_3}\\[.2cm] \nonumber
&=&-C\int d^dk\frac{1}{k^2-m^2} + 2 C (q^2+m^2)I_2(q^2,m^2)\,.
\eea
Therefore 
\be
\frac{d\Sigma_{\rm S}}{dq^2}{\Bigg\vert}_{q^2=m^2}\, =\, 2 C \left[
I_2(m^2;m^2) + 2 I_2^{\,'}(m^2;m^2)\right]\,,
\ee
and 
\be
\frac{1}{2}
\frac{d\Sigma_{\rm S}}{dq^2}{\Bigg\vert}_{q^2=m^2}\, =\,
C Q_{m^2}
\left( T_0^{\rm UV} - T_0^{\rm IR}\right)\,=\,0\,,
\ee
where 
\be
C Q_{m^2}=C_F\frac{\as}{4\pi}
\left(\frac{4\pi\mu^2}{m^2}\right)^{\ep}\Gamma(1+\ep) \,.
\ee
The self-energy of the massive scalar is zero. As in the case
of the massless fermion correction, the ultraviolet and infra-red
divergences exactly  cancel. The ultraviolet counter term is
non-zero
\be\label{SUVCT}
\left( \frac{1}{2} \Sigma_{\rm S}^{\,'} \right)^{\rm CT}\,=\,
C_F\frac{\as}{4\pi}
\left(\frac{4\pi\mu^2}{m^2}\right)^{\ep}\Gamma(1+\ep)
\left(-\frac{1}{\ep} + \ln \frac{\mu^2}{m^2}\right) \,.
\ee
Similarly for the massless fermion line we obtain
\bea 
\Sigma_{\rm F}&=&C(2-d)\int d^dk\frac{ {\rlap{p}/}_1-{\rlap{k}/} }{D_2 D_3}
\,=\,-2C(d/2-1)\frac{1}{2}\rlap{p}/_1\, I_2(p_1^2;0)\\[.2cm]\nonumber
&=&C_F\frac{\as}{4\pi}\frac{(4\pi)^{\ep}}{\Gamma(1-\ep)}
\left(-T_0^{\rm UV} + T_0^{\rm IR}\right)\,,
\eea
and therefore 
\be
\frac{1}{2}
\frac{ d\Sigma_{\rm F} }{ d{\rlap{p}/}_1 }{\Bigg\vert}_{{\rlap{p}/}_1=0}\, =\,
C_F\frac{\as}{4\pi}\frac{(4\pi)^{\ep}}{\Gamma(1-\ep)}
\left(-\frac{1}{2} T_0^{\rm UV} + \frac{1}{2} T_0^{\rm IR}\right)\,=\,0\,.
\ee
In dimensional
regularization, the self-energy of the massless fermion is zero: 
the ultraviolet and infra-red
divergences exactly cancel. 
Using the approximate relation (\ref{gam+gam-})
the ultraviolet counter term can be written  as
\be\label{FUVCT}
\left( \frac{1}{2} \Sigma_{\rm F}^{\,'} \right)^{\rm CT}\,=\,
C_F\frac{\as}{4\pi}
\left(\frac{4\pi\mu^2}{m^2}\right)^{\ep}\Gamma(1+\ep)
\left(\frac{1}{2\ep} - \frac{1}{2}\ln \frac{\mu^2}{m^2}\right) \,.
\ee
\subsubsection{Vertex corrections}
\noindent
The numerator of the vertex correction is given by
\bea
\overline{u}(p_2)
(-\rlap{k}/ + {\rlap{p}/}_1)(-\rlap{k}/ + 2 {\rlap{p}/}_2)u(p_1)
 &=&
\overline{u}(p_2)\left[k^2-2p_1k-4p_2k +
4p_1p_2\right]u(p_1)\\[.2cm]\nonumber\qquad
&=& T_{\rm B}\left(k^2-4kq+2p_1k+2m^2\right)\\[.2cm]\nonumber
 &=& T_{\rm B}
\left(2D_1-D_2+2m^2\right) \,.
\eea
The  vertex correction  can then be written as
\bea
V&=&T_{\rm B}C \int d^d k\frac{2D_1-D_2+2m^2}{D_1D_2D_3}
\\[0.2cm]\nonumber
&=&2m^2 C I_3(0,m^2;m^2) + 2 C I_2(p_2^2) - C I_2(m^2;m^2)\\[0.2cm]\nonumber
&=&C Q_{m^2}\left(-\frac{1}{\ep^2}-\frac{1}{\ep}-2 +{\cal O}(\ep)\right)
\eea
Ultraviolet divergences appear in $I_2(p_2^2)$ and $I_2(m^2,m^2)$, 
and therefore the counter term is
\be
V^{\rm CT}=T_{\rm B}
C_F\frac{\as}{4\pi}
\left(\frac{4\pi\mu^2}{m^2}\right)^{\ep}\Gamma(1+\ep)
\left(-\frac{1}{\ep} + \ln \frac{\mu^2}{m^2}\right) \,.
\ee
\subsubsection{Contribution to the subprocess cross section}
\noindent
The total contribution of the virtual corrections to the subprocess
cross section is
\bea
\hat\sigma^{\rm virt}&=& 2\hat\sigma_0 \left[
\frac{1}{2}\Sigma^{\,'}_{\rm S}+\frac{1}{2}\Sigma^{\,'}_{\rm F}+
V/T_{\rm B} + 
\left(\frac{1}{2}\Sigma^{\,'}_{\rm S}+\frac{1}{2}\Sigma^{\,'}_{\rm F}+
V/T_{\rm B}\right)^{\rm CT}\right]\\[.2cm]\nonumber
&=& \hat\sigma_0 
C_F\frac{\as}{2\pi}
\left(\frac{4\pi\mu^2}{m^2}\right)^{\ep}\Gamma(1+\ep)
\left(-\frac{1}{\ep^2} -\frac{5}{2\ep} -2
+ \frac{3}{2}\ln \frac{\mu_{\rm UV}^2}{m^2}\right) \,,
\eea
which leads to the final result
\be
\sigma^{\rm virt}\,=\, 
C_F\frac{\lambda^2\pi}{4\hat{s}}
\frac{\as}{2\pi}
\left(\frac{4\pi\mu^2}{m^2}\right)^{\ep}
\frac{1}{\Gamma(1-\ep)} {K}^{({\rm v})}_q \,,
\ee
where
\be
{K}^{({\rm v})}_q = \delta(1-z) 
\left(-\frac{1}{\ep^2} -\frac{5}{2\ep} -2-\frac{\pi^2}{6}
+ \frac{3}{2}\ln \frac{\mu_{\rm UV}^2}{m^2}+{\cal O}(\ep)\right) \,.
\ee

\subsection{Real contributions}
\subsubsection{Kinematics, matrix elements, counter terms}
We consider  the processes 
\bea\label{proc1}
e^+(p_2) + q(p_1) \rightarrow g(k) + \lq(q) \,,\\*[.2cm]\label{proc2}
e^+(p_2) + g(p_1) \rightarrow q(k) + \lq(q)\,.
\eea
The kinematical variables are 
\be
q^2=m^2\,,\quad \hat{s}=2p_1p_2=m^2+2kq\,,\quad
 \hat{t}=-2kp_1\,,\quad \hat{u}=-2kp_2\,.
\ee
The spin- and colour-averaged matrix element squared
of  process (\ref{proc1}) is given by
\be
\overline{\abs{T_{q}}}^2=\frac{\lambda^2}{4}C_Fg_s^2
\psi(p_2,p_1;k,q)_q \quad {\rm and} \quad \hat{\sigma}^{(0)}(\hat{s})_q=
\frac{\lambda^2}{4}C_Fg_s^2 R \left[\psi(p_2,p_1;k,q)_q\right]
\ee 
where  $R[\;]$ denotes the phase-space integral times 
the flux factor $\mu^{2\ep}/(2\hat{s})$. From explicit calculation
of the two Feynman diagrams we obtain  
\bea\label{psidotq}
\psi(p_2,p_1;k,q)_q&=&
\frac{m^4}{(kq) (kp_1)}-\frac{m^4}{(kq)^2}+
\frac{2m^2}{kp_1}+\frac{2kq}{kp_1}\\[.2cm]\nonumber
& & -\frac{2m^2}{kq}-2 +2\ep-2\ep\frac{kq}{kp_1}
\eea
Similarly, for the  process (\ref{proc2})  we obtain
\be
\overline{\abs{T_{g}}}^2=\frac{\lambda^2}{4}C_Fg_s^2
\psi(p_2,p_1;k,q)_g \,.
\ee 
where due to crossing symmetry 
\be\label{psidotg}
\psi_2(p_2,p_1;k,q)_g=-\frac{1}{1-\ep}\psi_2(p_2,-k;-p_1,q)_q \,.
\ee
Note that here we take  into account the fact
 that the spin average factor for an
initial gluon is $2(1-\ep)$.
The subprocess cross section is obtained by adding the 
$\overline{\rm MS}$ counter term
\be\label{CTxsec}
\sigma^{\rm CT}_a=
\frac{\as}{2\pi}(4\pi)^{\ep}\frac{1}{\ep\Gamma(1-\ep)}
\int_0^1 dz \; P_{b/a}(z,0)\; \hat{\sigma}^{\rm B}_b(z\hat{s})\,,
\ee
where $P_{b/a}(z,\ep)$ denotes the $\ep$ dependent and  spin-independent
Altarelli-Parisi splitting functions. In the present context we need
\bea
P_{q/q}(z,\ep)&=&C_F\left[\frac{1+z^2}{(1-z)_+}-\ep(1-z)+\frac{3}{2}\delta(1-z)
\right]\,,\\[.2cm]
P_{q/g}(z,\ep)&=&\frac{T_R}{1-\ep}\left[z^2 + (1-z)^2 -\ep\right]\,.
\eea
To perform the phase-space integrals it is convenient
to introduce  energy-angle variables in the subprocess 
centre-of-mass frame:
\bea
p_1^{\mu}&=&\frac{\sqrt{\hat{s}}}{2}(1,0,0,1)\,,\quad
p_2^{\mu}=\frac{\sqrt{\hat{s}}}{2}(1,0,0,-1)\\ \nonumber
k&=&\frac{\sqrt{\hat{s}}}{2}(1-z)(1,0,\sin \theta, \cos \theta)\,,\quad
q=p_1+p_2-k\,.
\eea
Changing variables to  
\be
z=m^2/\hat{s}\quad {\rm and}\quad u=\frac{1}{2}(1-\cos\theta)\,,
\ee
gives, for the  $d$-dimensional  phase space integral
 times flux factor,
\be
R[\;]=\frac{1}{32\pi\hat{s}}
\left(\frac{4\pi\mu^2}{\hat{s}^2}\right)^{\ep}\frac{1}{\Gamma(1-\ep)}
\int_0^1 du\, 2\, u^{-\ep}(1-u)^{-\ep}(1-z)^{1-2\ep}\,.
\ee
In terms of the 
variables $z$ and $u$ the Lorentz invariant scalar  products are
\be\label{ktperp}
m^2=z\hat{s}\,,\quad kp_1=\frac{\hat{s}}{2} u (1-z)\,,
\quad kq=\frac{s}{2}(1-z)\quad {\rm and} \quad 
 k_{T}^2=\hat{s}(1-z)^2u(1-u)\,.
\ee
\subsubsection{Contribution to $\hat{\sigma_1}$ from the $e^+q\to g\lq$ process}
In terms of the cms variables, the $\psi_q$ function (see
eq.~(\ref{psidotq}) becomes
\be
\psi(p_1,p_2;k,q)_q=4\left\{
\frac{1}{2u(1-z)}
\left[\frac{1+z^2}{1-z}-\ep (1-z)\right]
-\frac{z}{(1-z)^2}-\frac{1}{2}+\frac{1}{2}\ep
\right\}
\ee
Introducing notations for   the integrals
\bea
I_{-1}&=&\int_0^1 du u^{-1-\ep}(1-u)^{-\ep}=-\frac{1}{\ep}+\ep
\frac{\pi^2}{6} + \ldots\\[.2cm]
I_0&=&\int_0^1 du u^{-\ep}(1-u)^{-\ep}=1+2\ep + \ldots
\eea
and for the  functions of $z$
\bea
V_{-1}&=&(1-z)^{-1-2\ep}=-\frac{1}{2\ep}\delta(1-z)+\frac{1}{(1-z)_+}
-2\ep\left(\frac{\ln (1-z)}{1-z}\right)_+\\*
V_0&=&1\,,\qquad V_1=(1-z)
\eea
then
\bea
R[\psi_q] &=& \frac{1}{8\pi\hat{s}}
\left(\frac{4\pi\hat{s}^2}{m^2}\right)^{\ep}\frac{1}{\Gamma(1-\ep)}
\Bigl[
I_{-1} V_{-1} (1+z^2)\\ \nonumber
&-&\ep I_{-1} V_1 - 2 z V_{-1} I_0 -I_0 V_1
\Bigr]
\eea
and the unsubtracted cross section becomes
\be
\hat{\sigma}_{1q}=
C_F\frac{\lambda^2\pi}{4\hat{s}}
\frac{\as}{2\pi}
\left(\frac{4\pi\mu^2}{\hat{s}^2}\right)^{\ep}\frac{1}{\Gamma(1-\ep)}
\tilde{K}_q\,,
\ee
where
\bea
\tilde{K}_q&=&\delta(1-z)\left[\frac{1}{\ep^2} + \frac{1}{\ep}
+2-\frac{\pi^2}{6}\right]\\ \nonumber
&& -\frac{1}{\ep}\frac{1+z^2}{(1-z)_+}
-\frac{2z}{(1-z)_+}
+2\left(\frac{\ln (1-z)}{1-z}\right)_+(1+z^2)\,.
\eea
The $\overline{\rm MS}$ counter term  cross section (see
eqs.~(\ref{sigma0},\ref{CTxsec})) for this process is
\bea
\sigma^{\rm CT}_{1q}&=&C_F\frac{\lambda^2\pi}{4\hat{s}}
\frac{\as}{2\pi}
\left(\frac{4\pi\mu^2}{\hat{s}}\right)^{\ep}\frac{1}{\Gamma(1-\ep)}
\Biggl[
\frac{1}{\ep}\frac{1+z^2}{(1-z)_+} \\[.2cm]\nonumber
&& + \frac{3}{2}\delta(1-z)
-\ln \frac{\mu^2}{\hat{s}}\frac{1}{C_F}P_{q/q}(z,0) \,.
\Biggr] 
\eea
Finally for the subtracted  partonic cross section we get
\be
\hat{\sigma}_{1q}^{\rm sub}=
C_F\frac{\lambda^2\pi}{4\hat{s}}
\frac{\as}{2\pi}
\left(\frac{4\pi\mu^2}{m^2}\right)^{\ep}\frac{1}{\Gamma(1-\ep)}
{K}^{({\rm r})}_q
\ee
where 
\bea
{K}^{({\rm r})}_q&=&\delta(1-z)
\left[
\frac{1}{\ep^2} + \frac{5}{2\ep}
+2-\frac{\pi^2}{6}-\frac{3}{2}\ln\frac{\mu^2_{\rm IR}}{\hat{s}}
\right] \\[.2cm] \nonumber
&&-\frac{2z}{(1-z)_+}
+2\left( \frac{\ln (1-z)}{1-z} \right)_+(1+z^2)
-\ln \frac{\mu^2_{\rm IR}}{\hat{s}}\frac{1+z^2}{(1-z)_+} \,. 
\eea
\subsubsection{Contribution to $\hat{\sigma_1}$ from the $e^+g\to q\lq$ process}
In terms of the cms variables, the $\psi_g$ function (see
eqs.~(\ref{psidotq},\ref{psidotg})) becomes
\be
\psi(p_2,p_1;k,q)=\frac{4}{1-\ep}\left[\frac{1}{2u(1-z)}\psi_1(z,u)+
\psi_2(z,u)\right]\,,
\ee
where
\bea
\psi_1(z,u)&=&1-2z+\frac{2z^2}{1-u(1-z)}-\ep-(1-\ep)u(1-z)\,,\\*[.2cm]
\psi_1(z,0)&=&z^2+(1-z)^2-\ep\,,\\*[.2cm]
\psi_2(z,u)&=&\frac{1}{2}+
\frac{z^2}{\left[1-u(1-z)\right]^2}-\frac{z}{1-u(1-z)}\,.
\eea
The phase-space integral over $\psi_g$ including the  flux factor can be written
as
\be
R[\psi_g] = \frac{1}{8\pi\hat{s}}
\left(\frac{4\pi\mu^2}{\hat{s}^2}\right)^{\ep}\frac{1}{\Gamma(1-\ep)}
\tilde{K}_g
\ee
where 
\bea
\tilde{K}_g&=&
\Biggl\{\int_0^1 du\; \Bigl[-\frac{1}{\ep}\delta(u)+\frac{1}{u_+}
\Bigr](1+\ep)[1-2\ep\ln (1-z)]\psi_1(z,u)\\[.2cm]\nonumber
& &+ 2(1-z)\int_0^1 du\; \psi_2(z,u)\Biggr\} \,,
\eea
which leads to
\bea
\tilde{K}_g&=&
\Biggl\{- \frac{1}{\ep}{\left[z^2+(1-z)^2\right]}
+2\ln (1-z)\left[z^2+(1-z)^2\right]\\[.2cm]\nonumber
&  &+ 4z-4z^2 + 2z \ln z-2z^2\ln^2 z
\Biggr\} \,.
\eea
In this case the counter term simply cancels the $1/\ep$ term and
introduces the usual $\ln \mu^2 $ factor of the $\overline{MS}$ prescription.
Finally  we get
\be
\hat{\sigma}_{1g}^{\rm sub}=
T_R\frac{\lambda^2\pi}{4\hat{s}}
\frac{\as}{2\pi}
\left(\frac{4\pi\mu^2}{m^2}\right)^{\ep}\frac{1}{\Gamma(1-\ep)}
{K}_g\,,
\ee
where 
\be\label{Kg}
{K}_g=-\ln\left(\frac{\mu^2_{IR}}{\hat{s}(1-z)^2}\right)
\left[z^2+(1-z)^2\right] +2z(1-z)(2+\ln z)\,.
\ee
\subsubsection{Finite subprocess  cross sections}
\noindent

The finite subprocess cross sections are obtained
by adding the virtual and real contributions, whereupon  the remaining
soft and collinear singularities cancel:
\bea
\hat{\sigma}_{1q} &=&
\frac{\lambda^2\pi}{4\hat{s}}\frac{\as}{2\pi}\; C_F K_q \\
\hat{\sigma}_{1g} &=&
\frac{\lambda^2\pi}{4\hat{s}}\frac{\as}{2\pi}\; T_R K_g 
\eea
where
\bea
K_q & = & K_q^{\rm (v)} + K_q^{\rm (r)} \nonumber \\
& = & \delta(1-z)
\left(-\frac{\pi^2}{3} + 
\frac{3}{2}\ln \frac{\mu^2_{UV}}{\mu^2_{IR}}\right)\nonumber \\
&&
-\frac{2z}{(1-z)_+}
+2\left(\frac{\ln (1-z)}{1-z}\right)_+(1+z^2)
-\ln \frac{\mu^2_{\rm IR}}{\hat{s}}\frac{1+z^2}{(1-z)_+} \,,
\eea
and $K_g$ is given in eq.~(\ref{Kg}).
Finally, combining these with the appropriate parton distribution functions
gives the $e^+p$ cross section of eq.~(\ref{signlo}).
 
\subsection{Calculation of  $\langle k_{T}^2\hat{\sigma}_{q,g}\rangle$}

The average leptoquark transverse momentum squared is 
defined by the weighted subprocess cross sections
(see eq.~(\ref{ktperp}))
\bea
 \langle k_{T}^2\hat{\sigma}_q\rangle&=&\frac{\lambda^2\pi}{4}C_F g_s^2 R[
(1-z)^2u(1-u)\psi_q(z,u)]\,,\\[.2cm]\nonumber
 \langle k_{T}^2\hat{\sigma}_g\rangle&=&\frac{\lambda^2\pi}{4}T_R g_s^2 R[
(1-z)^2u(1-u)\psi_g(z,u)]\,.
\eea 
The integrations are finite in four dimensions and can be trivially
performed. One obtains
\bea
 \langle k_{T}^2\hat{\sigma}_q\rangle&=&\frac{\lambda^2\pi}{4}C_F\frac{\as}{2\pi} 
\frac{1}{3}\left[(1+z^2)(1-z)\right]\,, \\[.2cm]\nonumber
 \langle k_{T}^2\hat{\sigma}_q\rangle&=&\frac{\lambda^2\pi}{4}T_R\frac{\as}{2\pi}  
\left[\frac{1}{2}-3z+\frac{z^2}{2}+2z^3-4z^2\ln z\right]\,.
\eea 
Folding with the parton distribution functions and dividing by the total
cross section gives the result given in 
eqs.~(\ref{eqkt},\ref{eqktq},\ref{eqktg}).

\goodbreak


\vfill
\newpage
\vfill
\clearpage

\newpage
\vfill
\clearpage

\newpage
\vfill
\clearpage

\end{document}